\documentclass[12pt,preprint]{aastex}

\shorttitle{Particle acceleration in solar flare}

\shortauthors{Bykov & Fleishman}

\begin{document}

\title{Particle acceleration by strong turbulence in solar flares:
theory of spectrum evolution}

\author{A.M. Bykov\altaffilmark{1}, \and G.D. Fleishman\altaffilmark{2,1}}


\altaffiltext{1}{Ioffe Institute for Physics and Technology, 194021 St.Petersburg,
 Russia}

\altaffiltext{2}{New Jersey Institute of Technology, Newark, NJ
07102}

\begin{abstract}
We propose a nonlinear self-consistent model of the turbulent
non-resonant particle acceleration in solar flares. We simulate
temporal evolution of the spectra of charged particles accelerated
by strong long-wavelength MHD turbulence taking into account back
reaction of the accelerated particles on the turbulence. The main
finding is that the nonlinear coupling of accelerated particles with
MHD turbulence results in prominent evolution of the spectra of
accelerated particles, which can be either soft-hard-soft or
soft-hard-harder depending on the particle injection efficiency.
Such evolution patterns are widely observed in hard X-ray and
gamma-ray emission from solar flares.
\end{abstract}

\keywords{Sun: flares---acceleration of
particles---turbulence---diffusion---shock waves---Sun: X-rays,
gamma rays}

A solar flare arises due to fast and spatially localized  strong
energy release and reveals itself in electromagnetic radiation and
particle flows. Details of the energy release (as well as energy
storage) are currently debated. Models based on the idea of magnetic
reconnection are the most popular at the moment \citep{Shibata},
while interesting alternative possibilities, like balooning
instability \citep{Shibasaki}, or circuit models \citep{Zaitsev} are
discussed as well.

A common feature of the solar flares (as well as other astrophysical
objects with strong energy release) is the production of nonthermal
accelerated particles. There are now ample evidence of particle
acceleration in flares 
 \citep{Mayer, Nature90,Cane,Chupp, Akimov}. Accelerated electrons
reveal themselves in a variety of nonthermal emissions observed from
radio range to gamma-rays. Even rather small number of accelerated
electrons arising under a weak acceleration process can be visible
when the electrons produce coherent radio emission (e.g.,
\citet{Benz_1986, PRL}). Nonthermal incoherent  emissions
(gyrosynchrotron and/or bremsstrahlung) are detectable when the
acceleration is strong enough, providing a considerable fraction of
background electrons to be accelerated. 

Observations made by Ramaty High Energy Solar Spectroscopic Imager
(RHESSI) for the recent years have provided us with new stringent
constraints on the acceleration mechanism(s) operating in flares.
\citet{Grigis_Benz_2004, Grigis_Benz_2005} investigated spectral
evolution during individual subpeaks of the impulsive hard X-ray
(HXR) emission and found each such subpeak to display a
soft-hard-soft (SHS) evolution; the property, which was earlier
established for the impulsive phase as a whole
\citep{Parks_Winckler_1969, Kane_Anderson_1970, Benz_1977,
Brown_Loran_1985, Lin_Schwartz_1987}. \citet{Grigis_Benz_2004,
Grigis_Benz_2005} concluded accordingly that the ability to
reproduce the SHS spectrum evolution of the accelerated particle
population must be an intrinsic property of the acceleration
mechanism involved; it is the observational property of the
acceleration mechanism that is addressed in this Letter.

A number of acceleration mechanisms and models have been proposed to
account for the particle acceleration in flares (see, for a review,
\citet{Aschw_2002}). Acceleration by DC electric fields, both
sub-Dreicer and super-Dreicer, has been considered \citep{Holman85,
Tsuneta85, HolmanBenka, Litvinenko96}. This process is able to
provide the energization of particles up to $100$~keV, thus, it can
be considered as a possible mechanism of pre-acceleration in flares.

Stochastic acceleration by turbulent waves is currently assumed to
provide the main acceleration in impulsive solar flares
\citep{Miller96, Miller97, Petrosian92, Petrosian94, Petrosian97,
Petrosian98}, while the classical diffusive shock acceleration is
believed to play a role in large-scale gradual events.
\citet{Miller_etal_1997} made a detailed comparison between various
acceleration scenarios and concluded that the stochastic
acceleration is intrinsically consistent with the observational
constraints on the acceleration time, highest particle energy, and
the total number of the accelerated particles. In this Letter we
demonstrate that the turbulent stochastic acceleration is also
naturally consistent with the SHS spectrum evolution of the
accelerated particles.

\citet{Grigis_Benz_2006} noticed that within the standard model of
the stochastic acceleration \citep{Miller96} the higher level of the
turbulence results in harder steady-state spectrum of accelerated
electrons and vice versa, which looks consistent with the observed
SHS evolution. It is, nevertheless, unclear if the real evolution of
the spectrum will represent the sequence of such steady-state
spectra or will behave differently. We note, however, that a
fraction of stronger events (typically, the proton reach events)
displays a different kind of the spectral evolution, namely,
soft-hard-harder (SHH)
\citep{Frost_Dennis_1971,Cliver_1986,Kiplinger_1995,Saldanha_etal_2008},
as well as the gradual phase of the impulsive events
\citep{Grigis_Benz_2008}; we return to this point later.

%
%

Another important point firmly established by RHESSI data analysis
\citep{Brown_etal_2007, Dennis_etal_2007, Hudson_Vilmer_2007} is
that a significant fraction (some tens of percent) of the released
energy goes into nonthermal accelerated particles. This conclusion
is also confirmed by the radio data. \citet{Bastian_etal_2007}
performed a calorimetry of the accelerated electron energy in a
solar flare. They analyzed the radio spectrum evolution of a dense
flare, when most of the accelerated electron energy was deposited
into the coronal loop (rather than into the chromosphere). This made
it possible to accurately measure the total energy deposited by the
accelerated electrons into the coronal thermal plasma, which turned
out to be as high as 30\% of the estimated magnetic energy of the
flaring loop. These findings imply that the back reaction of the
acceleration particles on the accelerating agent (e.g., the
turbulence) is not negligible, in full agreement with time-dependent
test particle analytical solutions \citep{Byk_Fl_1992} and numerical
modeling \citep{Cargill_Vlahos_2006}, so this back reaction  must be
properly taken into account by the acceleration model.

As we have noticed, the stochastic acceleration of the charged
particles is the most promising candidate for the particle
acceleration in flares \citep[e.g.,][]{Miller_etal_1997,
Grigis_Benz_2006}. However, the mentioned energetic requirements
complicate strongly the whole theory of the stochastic acceleration.
First of all, the turbulence energy must be large enough to supply
the accelerated particles with sufficient energy. This means that
the turbulence is strong, so nonlinear effects are important
\citep[e.g.,][]{Yan_etal_2008} and the (quasilinear) approximation
of the weak resonant wave-particle interaction is invalid any
longer. Therefore, the theory must include a more general transport
equation valid in case of strong large-scale turbulence. Second,
since this turbulence loses a large fraction of its energy to
accelerate particles, this damping rate must be properly taken into
account; thus, one needs to solve two coupled equations -- one for
the particles and the other for the turbulence.

The renormalized theory of particle acceleration by strong
turbulence was developed by \citet{Bykov90a, Bykov90b}, see the
review by \citet{Bykov93} as well. The form of the equation for the
averaged distribution function of nonthermal particles depends on
whether the turbulence is composed of smooth (wave-like)
fluctuations only or contains also the shock fronts and other
discontinuities \citep{Bykov93}. Observations of the magnetic field
complexity in the flare-productive active regions
\citep{Abramenko_2005} along with  various models of primary energy
release in flares \citep{Shibata, Shibasaki, Zaitsev, Vlahos_2007}
suggest many ways of producing long-wave MHD turbulence
\citep[e.g.,][]{Miller_etal_1997, Grigis_Benz_2006} including
possibly a shock wave ensemble \citep{Anastasiadis_Vlahos_1991}.
Although the generation of discontinuities is likely under the
impulsive energy release \citep[e.g.,][]{Vlahos_2007}, we limit our
consideration to the case of smooth long-wave turbulence only, but
will include effect of the shocks in a further study.


%

We, therefore, adopt the following scenario. The process of flare
energy release is accompanied by formation of large scale flows and
broad spectra of MHD fluctuations in a reasonably tenuous plasma
with frozen-in magnetic fields. Vortex electric fields generated by
the compressible component of the large scale motions of highly
conductive plasma will result in  efficient \emph{non-resonant}
acceleration of charged particles.

The distribution function $N({\mbox{\boldmath $r$}},p,t)$ of
nonthermal particles averaged over an ensemble of turbulent motions
satisfies the kinetic equation
\begin{equation}\label{ke}
      \frac{\partial N}{\partial t} -
       \frac{\partial}{\partial r_{\alpha}} \: \chi_{\alpha \beta} \:
       \frac{\partial N}{\partial r_{\beta}}  =
      \frac{1}{p^2} \: \frac{\partial}{\partial p} \: p^4 D(t) \:
      \frac{\partial N}{\partial p} + F_{i}(p),
\end{equation}
The particle source term $ F_{i}(p)$ is determined by injection of
the electrons and nuclei. Although we do not consider explicitly the
injection process, we note that there are many ways to inject
particles into the stochastic acceleration by strong turbulence. The
possibilities are various versions of the DC acceleration
\citep{Litvinenko96, Litvinenko_2000, Litvinenko_2003b} or resonant
stochastic acceleration by the small-scale waves
\citep{Miller_etal_1997}. A nice option proposed recently by
\citet{FH08} is that the reconfiguration of the preflare magnetic
field can result in large-scale pulses of the Alfv\'en waves, which
in the presence of strong spatial gradients will generate
field-aligned electric regions capable of accelerating electrons
from the thermal pool up to 10 keV or above. Our further analysis
does not depend on the specific injection mechanism and details of
the particle injection process. Then, the phase space diffusion
coefficients $D$ and $\chi_{\alpha \beta} = \chi \,\delta_{\alpha
\beta}$ are expressed in terms of the spectral functions that
describe correlations between large scale turbulent motions (see
Bykov \& Toptygin, 1993). The kinetic coefficients satisfy the
following renormalization equations:

\begin{equation}\label{chi}
  \chi = \kappa + {5 \over 2} \int { d^3 {\bf k} \, d\omega \over (2\pi)^4 }
  \left[ {2T+S \over i\omega+k^2\chi }
        -{2k^2\chi S \over \left( i\omega + k^2\chi\right)^2 } \right] \, ,
\end{equation}

\begin{equation}\label{D}
  D={\chi \over 9} \int { d^3 {\bf k} \,  d\omega \over (2\pi)^4 }\;
   {k^4 S(k,\omega,t) \over \omega^2 + k^4 \chi^2 } \, ,
\end{equation}
where  $T(k,\omega,t)$ and $S(k,\omega,t)$  are the transverse and
longitudinal parts of the Fourier components of the turbulent
velocity correlation tensor. The equations for $T(k,\omega,t)$ and
$S(k,\omega,t)$ can be found in \citet{Bykov01}. We reproduce here
only the equation for the longitudinal spectral function
$S(k,\omega,t)$ responsible for the particle acceleration:

\begin{equation}\label{S_eq}
    \frac{\partial S(k,\omega,t)}{\partial t} -
       \frac{\partial \Pi_{\alpha}^S(k,\omega,t)}{\partial k_{\alpha}}  =
        \gamma_{ST}T(k,\omega,t) - \gamma_{dS} S(k,\omega,t) - \gamma_{ap}
        S(k,\omega,t).
\end{equation}

This full equation includes nonlinear cascading flux,
$\Pi_{\alpha}^S(k,\omega,t)$, as well as coupling with the
transverse function (term $\gamma_{ST}T(k,\omega,t)$), the true
damping (term $- \gamma_{dS} S(k,\omega,t)$, and the damping due to
acceleration of the charged particles (term $- \gamma_{ap}
S(k,\omega,t)$). Although all corresponding processes are generally
relevant for the turbulence evolution, we found that only the last
of them is critically important to provide the SHS spectrum
evolution due to the nonlinear nonresonant particle acceleration by
strong turbulence.  In case of a single scale long-wavelength
injection of the turbulent motions (gaussian spectrum with the
characteristic wave-number $k_0$) we can neglect both cascading term
in the left hand side and direct turbulence damping $\gamma_{dS}$ =
0. The turbulence is assumed to be confined in the acceleration
region;  possible turbulence leakage from the acceleration region is
compensated by the adopted sustained source of the transverse
component of the large-scale turbulence. Particles, however, can
escape from the region through its boundaries because of a large
mean free path of the particles outside the region. We, therefore,
consider a simplified equation for $S(k,\omega,t)$

\begin{equation}\label{S_eq_simple}
    \frac{\partial S(k,\omega,t)}{\partial t}   =
        \gamma_{ST}T(k,\omega,t) - \gamma_{ap}S(k,\omega,t),
\end{equation}
where the expression for the  damping rate of large scale turbulence
due to particle acceleration $\gamma_{ap}= \theta D$. The parameter
$\theta$ was determined (iteratively) in such a way to preserve
conservation of the total energy in the system of the turbulence and
the particles with account for energetic particle escape from the
acceleration region.  The standard Crank-Nikolson scheme providing
second order accuracy in time evolution was applied to integrate
Eqns. (\ref{ke})--(\ref{S_eq_simple}). The spatial transport term in
Eq. (\ref{ke}) was approximated as (time depending) escape time
$N/T_{esc}$, where $T_{esc} = R^2/4 \chi$,  $R$ is the
characteristic size of the acceleration region. Actually, the
temporal evolution of $\chi$ is very slow since in Eq. (\ref{chi})
it is dominated  by the transverse component of the turbulence
$T(k,\omega,t)$, which decays slower than the longitudinal component
$S(k,\omega,t)$.

The  temporal evolution of the turbulence is, thus, solely due to
particle acceleration effect. Particle acceleration time in the
model is longer than turn-over time of large scale MHD motions
providing $\omega > D$. Our simplified model assumes, therefore,
that the turbulence is primary produced in the form of transverse
motions with the scale about $2\pi/k_0$ with a gaussian spectrum,
which produces the corresponding longitudinal turbulence due to the
mode coupling (term $\gamma_{ST}T(k,\omega,t)$)
in a system of a finite scale size $R$ (where $R\, k_{\rm 0}
>1$). Furthermore the model accounts only for the evolution of large scale (energy containing) motions
of $k \, \lambda(p) \ll 1$, where $\lambda(p)$ is particle mean free
path due to scattering by small scale (resonant and non-resonant)
magnetic field fluctuations; the corresponding diffusion coefficient
is $\kappa = v \lambda(p)/3$. We did not consider here the
turbulence cascade to resonant (small) scales
\citep[c.f.][]{Miller96, PB08}. Instead, we fixed the "microscopic"
diffusion coefficient $\kappa(p)$ due to small scale fluctuations,
provided perhaps  by the whistler waves,  and considered the case of
intensive large scale turbulent motions provided $\kappa(p) < k_{\rm
0}^{-1}\cdot \sqrt{<u^2>}$.  The kinetics of particles satisfying
this inequality is determined by turbulent advection and so does not
depend on the details of the "microscopic" diffusion coefficient.

The energy range, where this inequality holds, does depend on the
charged particle mean free path $\lambda(p)$, which is ultimately
defined by generally unknown level of the small-scale turbulence.
\citet{Bastian_etal_2007} determined the mean free path of the radio
emitting electrons (a few MeV) to be about $10^7$ cm from the
characteristic decay time of the radio light curves for the case,
when the turbulent transport of the particles was independently
confirmed. This estimate is consistent with the small-scale (tens of
cm) turbulence level of about $\sim 10^{-7}-10^{-5}$ derived from
the decimetric continuum burst analysis \citep{Nita_etal_2005}.
Thus, to be conservative, in the case of electrons our approximation
is firmly justified at least up  to the energy about 1 MeV, where
the particle transport is fully driven by the large-scale turbulence
and as so it does not depend on the actual momentum dependence of
the mean free path $\lambda(p)$. Then the model accounts for a
nonlinear backreaction of accelerated particles on large scale
motions only. Although the assumed presence of the small-scale
turbulence implies the possibility of the stochastic resonant
acceleration along with considered here non-resonant acceleration,
we do not take into account the resonant acceleration explicitly,
because the energy density of the small-scale turbulence is much
smaller than that of large-scale turbulence. As has been discussed,
the resonant acceleration can, nevertheless, contribute to the
injection term $F_{i}(p)$ in Eq. (\ref{ke}).

\begin{figure}[tbp]
\plotone{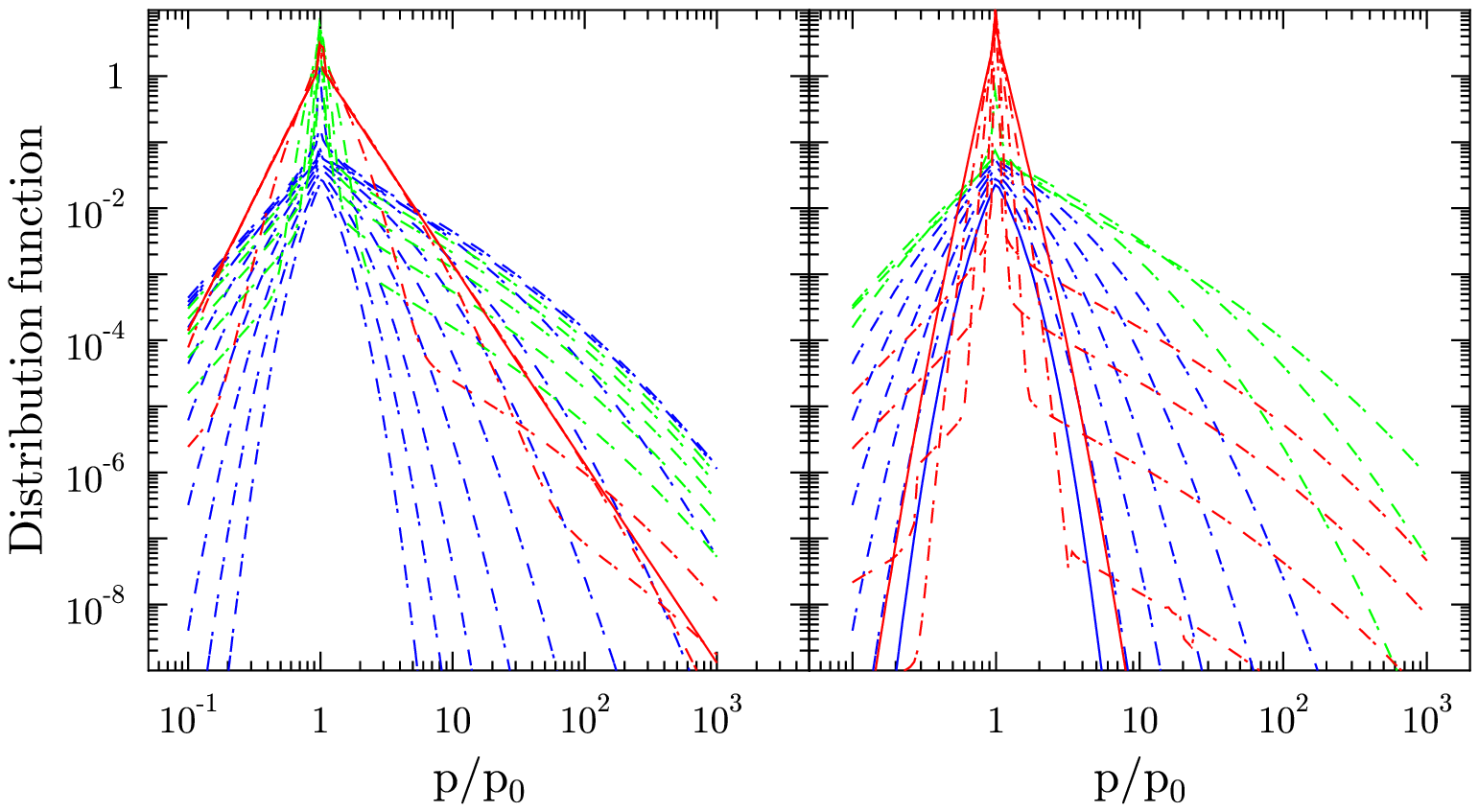} \caption{The temporal evolution of particle
distribution function (sequence of $p^2\, N$ vs $p/p_0$ plots, where
$p/p_0$ is the dimensionless particle momentum normalized by the
injection momentum $p_0$) simulated within a flare acceleration
region of the scale size $R = 14 \pi/ k_{\rm 0}$ for the particle
injection energy loading parameters $\zeta_{\rm i}$ = 10$^{-3}$
(left panel) and $\zeta_{\rm i}$ = 0.1 (right panel). Particle
spectra are shown in 20 logarithmically distributed consequent time
frames measured in $t\, D(0)$ starting from 0.01 to 30. For some
typical parameters, e.g., $R=2\times10^9$ cm, $B$=300 G, $n=
10^9-10^{11}$ cm$^{-3}$ we have $v_A\simeq 2.2\times (10^8-10^9)$
cm/s, and the characteristic acceleration time
$\tau_{acc}\equiv1/D(0)\simeq$ 1--10 s in agreement with HXR
\citep{Grigis_Benz_2006} and radio \citep{Bastian_etal_2007}
observations.\label{fig1}}
\end{figure}

We consider injection of non-relativistic particles of a momentum
$p_0$, i.e. $F_{\rm i}(p) \propto \delta(p - p_0)$. It is convenient
to characterize the injection efficiency by the injection energy
loading parameter
\begin{equation} \label{load}
\zeta_{\rm i} = \frac{2\, \int \epsilon(p) F_{\rm i}(p) p^2
dp}{D(0)\,\, \rho\, <u^2>},
\end{equation}
where $\epsilon(p)$ is the particle kinetic energy expressed via its
momentum $p$. In Figure~\ref{fig1} we show the particle distribution
function (normalized $\propto p^2~N$) calculated for the model.  We
assumed continuous injection of mono energetic particles (electrons
and protons ${\rm i} = {\rm e}, {\rm p})$ with the injection energy
loading parameters $\zeta_{\rm e}$ = 10$^{-3}$ (left panel) and
$\zeta_{\rm e}$ = 0.1 (right panel).


Although there are apparent differences in particle spectra for
different $\zeta_{\rm e}$, all our runs display clearly
soft-hard-soft behavior of the spectra of accelerated particles. The
origin of this spectral evolution is easy to understand within the
proposed model. Initial phase of the acceleration occurs in the
linear regime (test-particle approximation is still valid on this
stage), which results in effective particle acceleration by the
longitudinal large scale turbulent motions and spectral hardening.
However, fast particles accumulate a considerable fraction of the
turbulent energy on this stage and start to exhaust the turbulence,
thus, the efficiency of the acceleration decreases, which first
affects higher energy particles resulting in the spectrum softening.

Another important point, which must be noticed from the figure, is
that the slope of the spectrum at the late decay phase (red solid
curves) depends strongly on the injection efficiency $\zeta_{\rm
e}$. In fact, the spectrum is much steeper in case of strong
injection. In practice, the spectrum in the right panel is so steep
that it is probably undistinguishable against background thermal
particle distribution. This means that the sequence of the
(dash-dotted) spectra of accelerated electrons in the right panel
will reveal itself as  SHH evolution of the HXR spectrum. This
conclusion is consistent with the fact that the SHH evolution is
observed in stronger events, where enhanced injection of the charged
particles (e.g., protons) is likely, and with a recent finding of
gradual transitions between SHS and SHH evolution fragments
\citep{Grigis_Benz_2008}, which requires a common acceleration
mechanism for both SHS and SHH evolution patterns, even though
additional spectral hardening in the gamma-ray range can occur due
to relativistic particle trapping in the coronal loops
\citep{Krucker_2008}.

Besides the general SHS evolution, we should note that in agreement
with previous studies of the stochastic acceleration
\citep{Miller_etal_1997, Grigis_Benz_2006} the spectra do not obey
power-laws exactly: break-up and break-down turning points are
evident from the plots. It should be noted here that since the
nonlinear effects were taken into account in the model the
distribution function calculated for mono-energetic injection will
not have any of the general properties of the Green function of a
linear system. Therefore, one can not  build the distribution
function in the nonlinear case using the superposition principle any
longer. Nevertheless, the initial stage of the particle acceleration
occurs in the linear regime if the loading parameter $\zeta_{\rm e}$
is smaller than unity. Thus, broadly speaking,  the general behavior
of particle spectra evolution as it is illustrated in
Figure~\ref{fig1} will hold for other relatively narrow (as the
first blue curves in Figure~\ref{fig1}) initial particle
distributions with a similar loading parameter defined by Eq.
(\ref{load}).

\begin{figure}[tbp]
\plotone{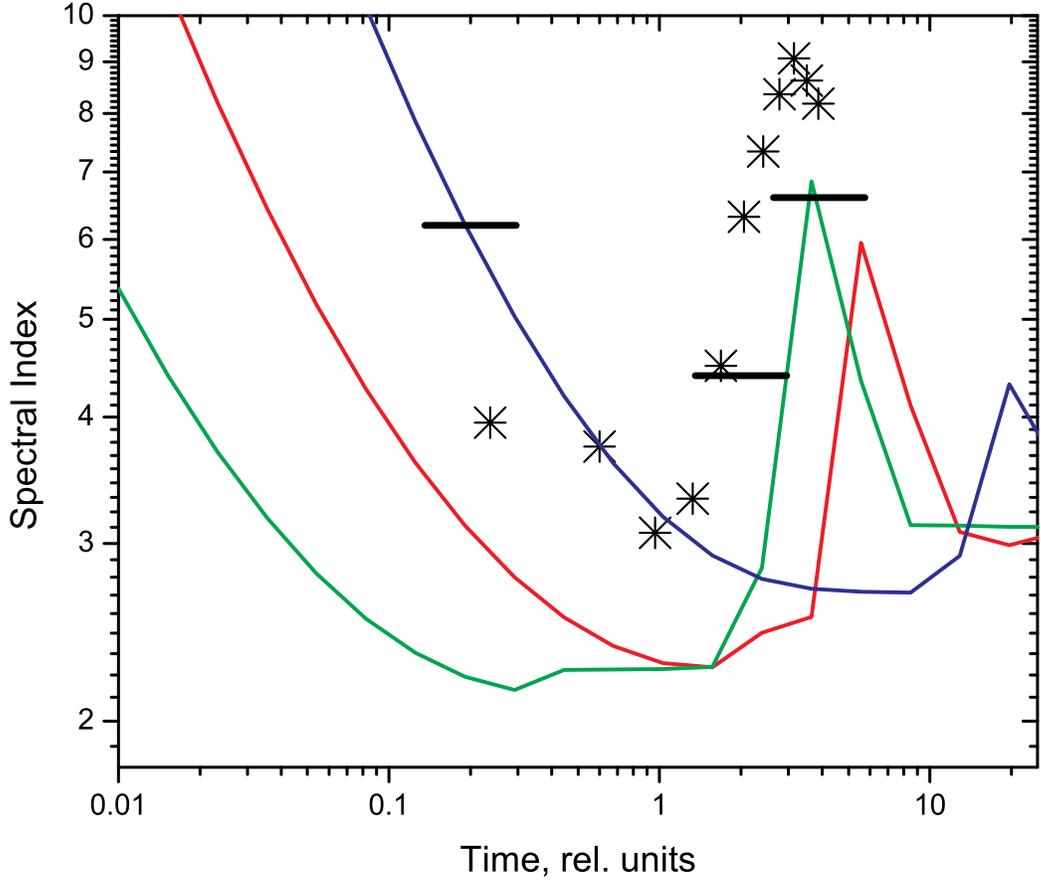} \caption{HXR spectral index evolution for
theoretically calculated spectra with various ratios of the escape
time to the acceleration time, $T_{esc}/\tau_{acc}=$ 5 (solid
curve), 1 (dashed curve), and 0.2 (dash-dotted curve); and observed
from the 04 Dec. 2002 flare, asterisks, \citep{Battaglia_Benz_2006},
and from the occulted 06 Sept. 2002 flare, horizontal dashes,
(E.Kontar, private communication) .\label{fig_HXR}}
\end{figure}

\citet{Grigis_Benz_2006} demonstrated that the HXR spectra from the
thin-target coronal sources are the most directly linked to the
energy spectra of the accelerated electrons, while the properties of
the thick-target foot-point sources can essentially be affected by
the transport effects. Accordingly, we computed the evolution of the
thin-target HXR emission generated by the evolving ensemble of the
accelerated electrons (as in Fig. 1) and then derived the evolution
of the HXR spectral index at $E=$ 35 keV to compare with the
observations of the coronal HXR sources reported by
\citet{Battaglia_Benz_2006}. The theoretical dependences of the HXR
spectral index on time are shown in Figure \ref{fig_HXR} by three
curves with different ratio of the acceleration time to the escape
time. The asterisks in the same plot show the evolution of the HXR
spectral index observed for the coronal source in the Dec., 04, 2002
event \citep{Battaglia_Benz_2006}. Even though no theoretical curve
is the exact fit to the data, one can clearly note important
similarities between theoretical and observational curves including
the main SHS behavior and some hardening at the later stage.

Since the spectral index analysis of the coronal source can in
principle be biased by much stronger foot-point contribution, a more
reliable way of the thin-target HXR analysis could be the study of
the occulted X-ray flares. However, the thin-target HXR flux is
typically weak from the occulted coronal sources, so the systematic
statistical study of the occulted flares reports only the spectral
data at around the peak time of the flare \citep{Krucker_Lin_2008}.
In some cases, however, it is still possible to derive information
on the spectral evolution of the occulted flares by integrating the
signal during rise, peak, and decay phases respectively. An example
of the corresponding spectral index evolution in an occulted 06
Sept. 2002 flare is shown by three long horizontal dashes in Figure
\ref{fig_HXR} (E. Kontar, private communication). The SHS evolution
is evident in this instance as well.

In addition to the mentioned similarities between the theoretical
and the observed spectra, there are also apparent differences. We
have to note, however, that the differences between the theory and
observations are less significant than the difference between the
spectra observed from different events. Thus, we can ascribe these
differences to the varying geometry of the source and/or to
different regimes of the turbulence generation, cascading, damping,
and escape, i.e., to those details of the model, which have not been
specifically addressed within this letter.



To summarize, we note that taking into account the nonlinearity,
which is necessarily present in a system where efficient
acceleration by strong turbulence occurs, offers a plausible way of
interpreting both kinds of the characteristic spectrum evolution,
SHS and SHH, observed from solar flares. A side achievement of the
adopted here model of the turbulent electron transport is the energy
independent escape time from the acceleration region, which implies
that electrons with different energies leave the acceleration site
simultaneously: the property required by measurements of the HXR
fine structure timing \citep{Aschw_2002}. A full comprehensive
picture of the particle acceleration in flares will require further
analysis with the shock waves, turbulence cascading, and injection
details included, as well as computing the HXR and gamma-ray
spectrum evolution, which we plan to consider elsewhere.

\acknowledgements

We are cordially grateful to Marina Battaglia for providing us with
the data on the spectral index evolution from the coronal sources,
and to Eduard Kontar, who obtained and provided us with the data on
the spectral evolution of the occulted 06 Sept. 2002 flare. The work
was partly supported by Russian Foundation for Basic Research,
grants No.06-02-16844, 06-02-16295, 08-02-92228, 09-02-00624 and by
NSF grants AST-0607544 and ATM-0707319 to New Jersey Institute of
Technology. We have made use of NASA's Astrophysics Data System
Abstract Service.

\bibliographystyle{apj}

\bibliography{acceler}

\clearpage

\end{document}